\documentclass[apj]{emulateapj}
\usepackage{natbib}
\usepackage{color}

\usepackage{graphicx}
\usepackage{wasysym}
\usepackage{amssymb}
\usepackage[flushleft]{threeparttable}
\newcommand{\unit}[1]{\ensuremath{\, \mathrm{#1}}}
\newcommand{\popiii}{{pop~{\sc iii}}}
\newcommand{\Popiii}{{Pop~{\sc iii}}}
\newcommand{\popii}{{pop~{\sc ii}}}

\newcommand{\HI}{{H~{\sc i}}}
\newcommand{\snia}{SN{\sc i}a}
\newcommand{\snii}{SN{\sc ii}}

\def \aj {Astron. J.}
\def \mnras {Mon. Not. R. Astron. Soc.}
\def \apj {Astrophys. J.}
\def \apjs {Astrophys. J. Suppl.}
\def \apjl {Astrophys. J. Let.}
\def \aap {Astron. \& Astrophys.}

\def \nat {Nature}

\def \araa {ARA\&A}

\shorttitle{How the first stars shaped the faintest gas-dominated dwarf galaxies}
\shortauthors{Verbeke et al.}

\begin{document}

\title{How the first stars shaped the faintest gas-dominated dwarf galaxies}
\author{R.~Verbeke$^{\dagger}$,  B.~Vandenbroucke,  S.~De Rijcke }
\affil{Sterrenkundig Observatorium, Ghent University, Krijgslaan 281, S9, 9000 Gent, Belgium}
\email{$^{\dagger}$robbert.verbeke@UGent.be}
\received{2015 June 10}
\revised{2015 October 23}
\accepted{2015 November 3}

\begin{abstract}

Low-mass dwarf galaxies are very sensitive test-beds for theories of
cosmic structure formation since their weak gravitational fields allow
the effects of the relevant physical processes to clearly stand
out. Up to now, no unified account exists of the sometimes seemingly
conflicting properties of the faintest isolated dwarfs in and around
the Local Group, such as Leo T and the recently discovered Leo~P and Pisces~A
systems. Using new numerical simulations, we show that this serious
challenge to our understanding of galaxy formation can be effectively
resolved by taking into account the regulating influence of the
ultraviolet radiation of the first population of stars on a dwarf's
star formation rate while otherwise staying within the standard cosmological
paradigm for structure formation. These
simulations produce faint, gas-dominated, star-forming dwarf galaxies
that lie on the baryonic Tully-Fisher relation and that successfully
reproduce a broad range of chemical, kinematical, and structural
observables of real late-type dwarf galaxies. Furthermore, we stress the importance
of obtaining properties of simulated galaxies in a manner as close as possible to
the typically employed observational techniques.
\newline

\end{abstract}


\section{INTRODUCTION}

Leo P \citep{giovanelli13} and Pisces A \citep{tollerud15} are the latest
additions to a growing list of faint, gas-rich,
isolated dwarf galaxies; a list that dates back at least to the
discovery of Leo~T \citep{irwin07}. These galaxies have luminosities of only a few $10^5~\unit{L_\odot}$ \citep{ryanweber08,mcquinn13,tollerud15} and {\HI} masses $3-4$ times higher than their stellar mass. 
Rotation velocities, derived from radio observations of the {\HI}, 
are estimated at $v_{\rm rot} \sim 12.5\unit{km~s^{-1}}$
\citep[Leo~T:][]{ryanweber08}, $v_{\rm rot} \sim 17\unit{km~s^{-1}}$
\citep[Pisces~A:][]{tollerud15} and   $v_{\rm rot} \sim 18\unit{km~s^{-1}}$
\citep[Leo~P:][]{giovanelli13,bernsteincooper14}. The available data suggest that all three
dwarfs formed stars continuously, although at a very low and highly
variable rate \citep{clementini12, weisz12, mcquinn13, mcquinn15}.

The existence of these systems poses a strong challenge to $\Lambda$CDM, the
standard model for galaxy formation and evolution: cosmological simulations
predict that half of the galaxies in the nearby Universe with a
circular velocity $\sim 25\unit{km~s^{-1}}$ are dark; these halos were never able
to form stars. At a circular velocity of $\sim 15\unit{km~s^{-1}}$, comparable to
Leo~T, Leo~P, and Pisces~A, over 90\% of all halos is predicted to be
dark \citep{sawala14}. Indeed, supernova explosions together with the
cosmic ultraviolet background (UVB), produced by the first galaxies
and quasars \citep{ef92, haardtmadau96, fauchergiguere09}, tend to
remove the star-forming gas from low-mass dwarf galaxies. However, \cite{tollerud15} found that an order of magnitude difference between the observed number density of {\HI}-detected faint dwarfs and that of corresponding dark-matter halos predicted from cosmological simulations is highly unlikely.

Clearly, these are not rare objects and many more such faint
systems are likely awaiting discovery \citep{adams13, faerman13,
  adams15, janesh15, sand15}. Thus, there appears to be a disagreement between
the predicted and the observed abundance of faint, gas-dominated,
star-forming dwarf galaxies near the Local Group.

The process of galaxy formation is very challenging to model and the observable properties of simulated galaxies will strongly 
depend on the chosen set of parameters. While parameters may be tuned to get one or several 
galaxy properties in agreement with observations, reproducing a broad range of them is not a 
trivial task. Failing to reproduce one or more observable properties may be indicative that an important astrophysical
process is not taken into account in the models. We thus argue that to truly reproduce realistic galaxies, one has to look at all the known observable galaxy properties. Many efforts have been made to self-consistently simulate 
the formation and evolution of low-mass dwarf galaxies \citep[e.g.][]{governato10, 
trujillogomez11, munshi13, cloetosselaer14, shen14, benitezllambay15, onorbe15, sawala15, trujillogomez15, wang15}. 
In this paper, we compare our simulations with as many observable properties as possible. 

Here, we propose a theoretically attractive way of alleviating these
tensions between simulations and observations: the energetic feedback
from the first stars that formed in the universe. These
so-called ``population {\sc iii}'', or \popiii, stars are expected to
have very different properties than stars born out of even very weakly
enriched gas, which are called ``population {\sc ii}'', or \popii,
stars. As we will show, this allows faint dwarfs with
stellar masses of $M_\star \sim 10^{5-6}\unit{M_\odot}$ to grow in
dark-matter halos of $M_{\rm DM} \sim 10^{9-10}\unit{M_\odot}$ which
are massive enough to retain their cold gas reservoirs.

Below, we discuss ten numerical simulations of dwarf galaxies with
different mass assembly histories in a cosmological setting with added
{\popiii} feedback. We also ran convergence test simulations and
control simulations with the same initial conditions and mass assembly histories but without
{\popiii} feedback. The simulated dwarfs cover the entire mass range of gas-rich dwarf galaxies, 
allowing for a comparison with Leo~T, Leo~P, and Pisces~A. We refer to Table
\ref{tab:simulations} for an overview of the simulations. In Section 2, 
we discuss the code, initial conditions and analysis methods we used. 
The results of our simulations are shown and discussed in Section 3. 
Section 4 provides a short summary of this work.

\section{SIMULATIONS}

Our simulations were performed with the N-body/SPH-code GADGET-2 \citep{springel05} to which we added several
astrophysical ingredients, including radiative cooling, heating by the
cosmic UVB and the interstellar radiation field, star formation,
supernova and stellar feedback and chemical enrichment. 

It is known that the standard SPH prescription suffers from several numerical issues \citep[see e.g.][and references therein]{springel10, hopkins15}. These are most notable in processes such as ram-pressure stripping. For our simulated galaxies, which do not experience processes where the standard SPH prescription gives significantly erroneous results, corrections to the hydrodynamical scheme will be unlikely to have a significant influence on the results.

\begin{table*}[t!]
\begin{minipage}{0.95\textwidth}
\begin{center}
\caption{Overview of the simulations.}

\label{tab:simulations}

\resizebox{\textwidth}{!}{
\begin{tabular}{lllllllcl}
\hline

(1) 		& (2) & (3) & (4) & (5) & (6) & (7) & (8) & (9) \\ 
Name		&	$M_\mathrm{DM, tot}$			&	$M_\mathrm{prog}$ & $M_\mathrm{res}$		&	$n_\mathrm{DM}$	& $n_\mathrm{SPH}$  &  $\epsilon_f$ &	Symbol &	Notes\\
		&	$[10^{9}\unit{M}_\odot]$	&	$[10^{9}\unit{M}_\odot]$ 	&	$[10^{7}\unit{M}_\odot]$	&	$\times 10^5$		& $\times 10^5$ & $[\unit{pc}]$ & &\\ \hline
DG9e9	&	9	&	1.36		&	5		&	4.5	&	18 	& 6.15 & $\star$			& \\	
DG10e9	&	10	&	1.82		&	5		&	5	&	20 	& 6.15 & $\Circle$			& \\	
DG12e9	&	12	&	1.88		&	5		&	6	&	18 	& 6.77 & $\triangleleft$ 			& \\	
DG13e9a	&	13	&	1.86		&	5		&	6.5	&	6.5	& 9.76 & 	$\bigtriangledown$ 		&  \\
DG13e9b	&	13	&	2.04	 	&	5		&	6.5	&	6.5	& 9.76 & 	$\bigtriangleup$ 		&  Different initial conditions\\
DG15e9a	&	15	&	1.82		&	5		&	7.5	&	7.5	& 9.76	& 	$\Square$ 		& \\
DG15e9b	&	15	&	1.57		&	5		&	7.5	&	7.5	& 9.76	& 	$\diamond$ 		& Different initial conditions\\
DG20e9	&	20	&	3.07 	&	10		&	10	&	10	& 9.76 	& $\pentagon$ 		&\\	
DG50e9  &	50	&	2.98		&	10		&	10	&	10	& 13.25	&	$\hexagon $			&	\\
DG1e11  &	100	&	7.34		&	20		&	20	&	20	& 13.25	&	$\varhexagon $			&	\\ \hline
DG15e9b-CT	& 15 	&	1.57 &	5	&	7.5	&	15	& 7.75 	& $\diamond$	& Convergence test of DG15e9b\\	\hline	
DG10e9-NP3	&	10	&	1.82		&	5	&	5	&	5 	& 9.76 & $\Circle$ 	& DG10e9 without \Popiii\ feedback\\	
DG12e9-NP3	&	12	&	1.88		&	5	&	6	&	6	& 9.76 	& $\triangleleft$	& DG12e9 without \Popiii\ feedback\\	
DG13e9b-NP3	&	15	&	2.04		&	5	&	7.5	&	7.5	& 9.76 & $\bigtriangleup$ 	& DG13e9b without \Popiii\ feedback\\	

\hline
\end{tabular}}
\end{center}
\begin{tablenotes} 
\item[] \textbf{Notes.} (1): the name of the simulation, (2): total dark matter mass in the simulation, (3): the dark matter mass of the most massive progenitor, (4): the minimal progenitors mass in the merger tree, (5): number of dark matter particles, (6): number of gas particles, (7): force resolution, (8): symbol used in the figures, (9): additional notes.
\end{tablenotes}
\end{minipage}

\end{table*}

\subsection{Astrophysical prescriptions}
As a model for star formation, a gas particle can be converted into
a star particle with the same metallicity, position, and velocity if
the local velocity field is converging and if the gas density is
sufficiently high. With a density threshold of $100\unit{amu~cm^{-3}}$, star
formation is restricted to cold, dense clouds of neutral gas. Each
star particle represents a stellar population following the initial-mass 
function (IMF) of \cite{chabrier03}. We include
thermal feedback by young, massive stars, supernovae of type {\sc i}a
(\snia), and type {\sc ii} supernova (\snii). Young O and B stars are
assumed to each inject $10^{50}$~ergs of thermal energy into the
ISM while {\snii} and {\snia} explosions produce
$10^{51}$~ergs per event, with an absorption efficiency of 70\% by 
the ISM \citep{cloetosselaer12}. The delay times for {\snia}
explosions are normally distributed with a mean of $\mu=4\unit{Gyr}$
and a dispersion $\sigma=0.8\unit{Gyr}$ \citep{strolger04}, cut off at
$3\sigma$.  \snii~and \snia~also enrich the ISM and, given these are the only two sources
of enrichment, we only need to explicitly follow the evolution of two elements (e.g. Mg and Fe) to know the full
chemical composition \citep{derijcke13}.
The radiative cooling and heating rate is
redshift, density, temperature and metallicity dependent, taking into account
self-shielding by neutral Hydrogen in high-density particles.
We take into account the effect that part of the energy injected into the gas
by the UVB is used to ionize the gas \citep{vandenbroucke13}. 
All these prescriptions and techniques
have been used and thoroughly tested previously and a more in-depth
discussion of them is available in the literature \citep[see][]{valcke08, schroyen11, cloetosselaer12, schroyen13, verbeke14,
  cloetosselaer14}.

\subsection{{\Popiii} stars}

\begin{figure}[t!]
\includegraphics[width=0.49\textwidth]{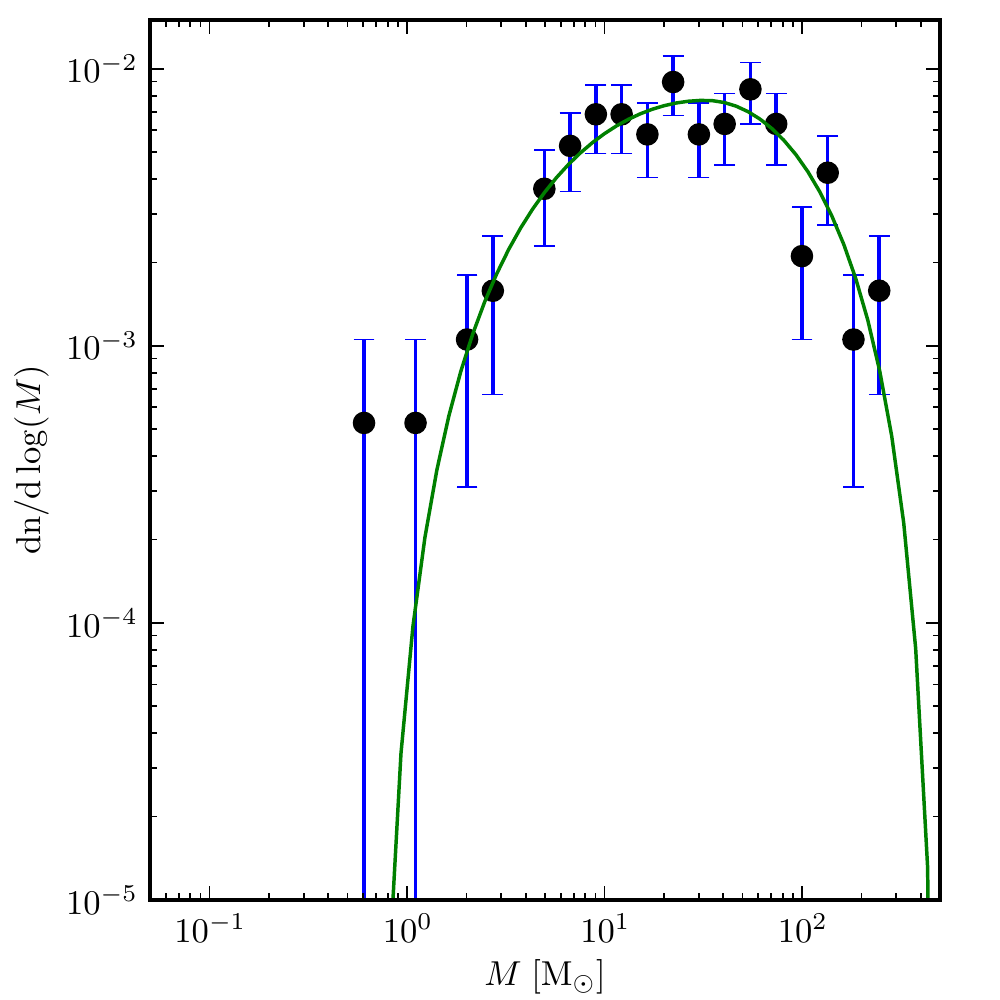}
\caption{Initial mass function. We use a fit (green line) through the data points (black dots) from \cite{susa14}.}
\label{fig:imf}
\end{figure}

\Popiii~ stars formed in the early Universe out of
pristine gas. To date no {\popiii} stars have been directly observed
and their properties are necessarily derived from theory. Simulations predict
{\popiii} star masses in the range between 0.7 and 300 $\unit{M_\odot}$,
with a significantly more top-heavy IMF than the Chabrier function of
{\popii} stars \citep[][Fig. \ref{fig:imf}]{susa14} . Compared with a {\popii} particle, a
{\popiii} star particle injects 4 times as much thermal energy into
the ISM by {\snii} explosions and 40 times as much by young massive
stars \citep{heger10}. 45\% of the mass of a
{\popiii} particle is returned to the ISM, with the remainder locked
up in remnants, while the chemical yields of Mg and Fe are taken to be
10\% of the enrichment by normal {\snii} explosions \citep{heger10,
  nomoto13}. The latter is an approximation but, given the current
theoretical uncertainties on {\popiii} yields and given the fact that
a negligible fraction of the elements eventually present in the ISM
comes from {\popiii} stars, this is without consequence. A star
particle is treated as a {\popiii} particle if $\mathrm{[Fe/H]} < -5$.

Recently, several other advanced treatment methods of rapid stellar feedback have been suggested and shown to be important in galaxy evolution models \citep[e.g.][]{stinson13, hopkins14}. The main difference between these and the {\Popiii} feedback proposed here is that the latter is inherently redshift dependent.

\subsection{Initial conditions and merger trees}

\begin{figure*}[t!]
\begin{minipage}{\textwidth}
\begin{center}
\includegraphics[width=\textwidth]{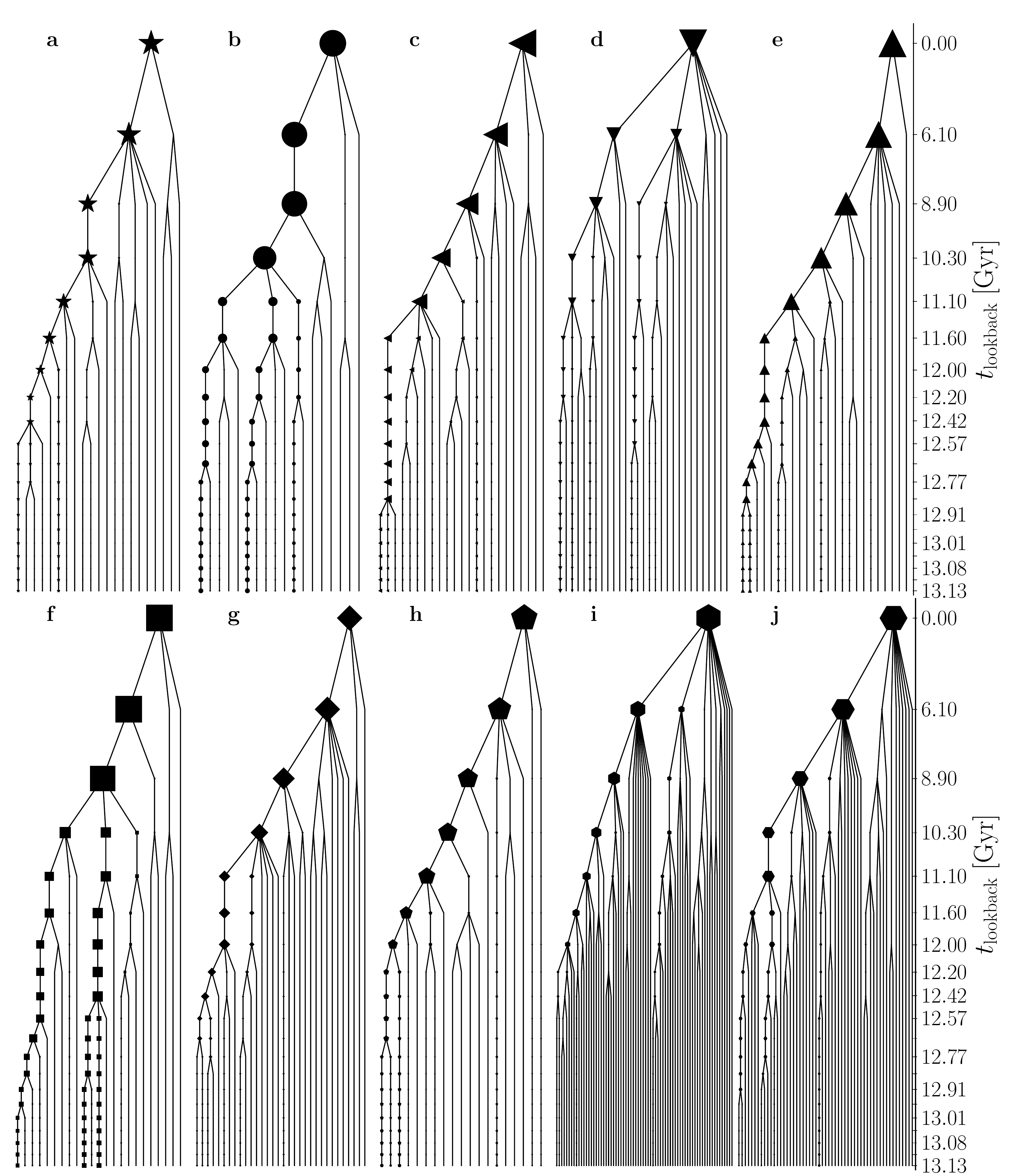}
\end{center}

\end{minipage}

\caption{Merger trees of the simulations. (a) DG9e9, (b) DG10e9, (c) DG12e9, (d) DG13e9L, (e)
DG13e9E, (f) DG15e9L, (g) DG15e9E, (h) DG20e9, (i) DG50e9 and (j) DG1e11, and their respective
re-simulations without \popiii~ stars and the convergence test. The size of the marker is
indicative of the total mass of a halo at the corresponding lookback time.
}

\label{fig:trees}
\end{figure*}

The mass assembly history of a simulated dwarf galaxy is modeled as a
merger tree constructed using the extended Press-Schechter formalism \citep{bond91, parkinson08} for a $\Lambda$CDM cosmology with the parameters 
$\Omega_{m} = 0.2726$, $\Omega_\Lambda = 0.7274$, $\Omega_\mathrm{bar} 
= 0.0476$ and $\mathrm{H_0} = 70.4  \unit{km~s^{-1}~Mpc^{-1}}$, which is consistent with the results from the 
WMAP-9 \citep{hinshaw13}. We simulate the 
progenitors of the dwarf galaxy initially in isolation starting from $z=13.5$, and subsequently let them coalesce
as prescribed by their merger tree, shown in Fig. \ref{fig:trees}. 
The progenitors are given an initial
rotation, with both the magnitude and direction of the initial angular
momentum selected randomly. 
Each merger event is treated as a two-body
interaction with orbital parameters drawn from probability
distribution functions derived from cosmological
simulations \citep{benson05}.The density
profile of each dark matter halo is given by:
\begin{equation}
\rho_\mathrm{DM} = \frac{\rho_0}{(r/r_s)^\alpha(1+r/r_s)^{4-2\alpha}},
\end{equation} 

with $r_s$ the scale length, defined as the radius where the logarithmic slope of the density profile is $-2$.
The values for $\alpha$ and $r_s$ are drawn from cosmologically motivated mass and redshift dependent 
probability distribution functions \citep{cen04}. Although similar generalizations of the NFW profile \citep{navarro96} exist \citep{dutton14, klypin14}, the profiles used here have the advantage of being derived for high redshifts and down to halo masses of $M \sim 10^{6.5} M_\odot$. Furthermore, the density profile is significantly influenced by stellar feedback and protogalaxy mergers, so we argue that the initial density profile has little influence on the galaxy properties at $z=0$. 
The initial density of the gas is given by a pseudo-isothermal profile \citep{revaz09}: 
\begin{equation}
\rho_g(r) = \frac{\rho_0}{1+r^2/r^2_g},
\end{equation}
with $\rho_0$ the central gas density and $r_g$ the scale length, determined in the same way as in \cite{schroyen13}. Initially, there are no stars present in the simulation.

We do not explicitly include accretion in the simulations, however, the simulated galaxies are surrounded by a very extended diffuse gas halo which does allow for smooth gas accretion implicitly in a self-consistent way. Similarly, the galaxies are embedded in an extended dark matter halo, which allows for growth of the dark matter halo through accretion.

\subsection{Resolution and convergence test}
All but two simulations were run with the mass of the dark matter
particles fixed at $m_\mathrm{DM} = 2\times 10^4\unit{M_\odot}$, except for
DG50e9 and DG1e11 with $m_\mathrm{DM} = 5\times 10^4~
M_\odot$. In most simulations, the mass of the gas particles was
simply scaled according to the assumed baryonic to dark-matter density 
ratio: $\Omega_\mathrm{bar}/\Omega_\mathrm{DM} = 0.2115$.
Exceptions to this are DG10e9, DG12e9 and DG15e9b-CT, in
which the gas resolution was increased respectively with a factor of 4, 3 and 2. 
In the latter, this was done as a convergence test to ensure that our results do 
not change depending on the resolution. As can be seen (Figs. \ref{fig:btf}$-$\ref{fig:properties}),
DG15e9b-CT does not differ significantly from DG15e9b and any changes
can be attributed to the stochastic nature of star formation, showing that
the models are sufficiently converged. In DG10e9 and DG12e9, the resolution was increased because of the low star
formation efficiency in low mass halos. 
The force resolution of each simulation is fixed by the mass of the gas particles and the star formation criteria: the softening length is the same as the smoothing length of a gas particle that satisfies the density threshold. These are given in column 7 in Table \ref{tab:simulations}. The softening length is the same for all the baryonic and dark matter particles.

\subsection{Analysis}

The post-processing analysis of the simulations was performed with our
own publicly available software package HYPLOT\footnote{http://sourceforge.net/projects/hyplot/}.
For this analysis, we adopted the strategy of staying as close as possible to the strategies
adopted by observational astronomers. The comparison between properties obtained in this manner 
and those obtained in a way that is typically done by theorists, will be presented in an upcoming
paper (Vandenbroucke et al. 2015, in preparation).
The absolute magnitudes were determined by fitting an exponential function to a
simulated galaxy's surface brightness profiles out to $R_{30}$, the
point where the surface brightness drops to $30 \unit{mag~arcsec^{-2}}$,
and extrapolating it to infinity to estimate the total luminosity. The
V-band half-light radius $R_e$ was determined from this fitted
profile. The stellar mass $M_\star$ is estimated from the V and I-band
luminosities \citep{bell01}. The neutral gas mass
$M_\mathrm{HI}$ is determined by a straightforward summation of the
masses of the gas particles multiplied with their neutral
fraction. The UVB ionizes the outer, low-density regions of the ISM,
limiting the {\HI} to the more central, dense parts of the ISM of a
simulated dwarf galaxy. The total baryonic mass is $M_\mathrm{bar} =
M_\star + M_\mathrm{gas, atomic}$, with the atomic gas mass corrected
for the Helium fraction: $M_\mathrm{gas, atomic} =
M_\mathrm{HI}/(1-\mathrm{Y}_\mathrm{He})$ (with the primordial Helium
abundance $\mathrm{Y}_\mathrm{He} = 0.25$). In order to produce an
``observational'' estimate for the circular velocity $v_c$, we fitted
a gaussian to a mock {\HI} spectrum of the galaxy viewed edge-on and adopted $v_c = W_{20}/2$,
with $W_{20}$ the full width at 20\% of the maximum of this gaussian. 
The cumulative star formation histories (CSFH) were
determined from the birth times and masses of the stellar particles
within a certain radius. These CSFHs, like the observed
ones, thus do not take mass loss by SNe into account. Using stellar
evolution tracks for \popii~ stars \citep{bertelli08, bertelli09} and
for \popiii~ stars \citep{marigo01}, the number of RGB stars within a
given star particle can be calculated. Thus, the mean Iron abundance,
$\mathrm{\langle[Fe/H]\rangle}$, can be calculated via a sum over all stellar
particles weighted by the number of RGB stars in each particle. This
mimics the procedure actually followed for real
dwarfs \citep{kirby13}. For the gas, we computed the Oxygen abundance of
dense gas, with each gas particle weighted by its ionized
fraction. This is a measure for the metallicity of the ionized gas
around star-forming regions. The SFR and atomic gas density, required
for the Kennicutt-Schmidt relation, was determined within an aperture
with a radius of $200\unit{pc}$ around the galaxy center. The 1D
stellar velocity dispersion $\sigma_\star$ is measured along a
line-of-sight through the galaxy, viewed edge-on, within an aperture of
1 half-light radii around the galaxy center. The solar logarithmic
mass-fraction abundance of Iron is assumed to be $\mathrm{Fe/H}_\odot
= -2.756$ and the Magnesium to Iron ratio is $\mathrm{Mg/Fe}_\odot =
-0.261$ \citep{asplund09}.  The virial radius $R_{{\rm halo}}$ of a dark matter
halo is the radius wherin the average density equals 200 times the mean
density of the universe ($\rho_\mathrm{Univ} = 9.47\times
10^{-30}\unit{g~cm^{-3}}$), while the virial mass $M_{{\rm halo}}$ is the total
dark matter mass within $R_{{\rm halo}}$: $M_{{\rm halo}} =\frac{4\pi}{3}
200\rho_\mathrm{Univ}R_{{\rm halo}}^3$.

\section{RESULTS}

In what follows, all simulated galaxy properties have been calculated
at the present epoch, unless explictly stated otherwise, to enable a
comparison with dwarf galaxies observed in the nearby Universe. Some 
of the properties of the simulations at $z=0$ are given in Table \ref{tab:properties}.

\begin{table*}[t!]
\begin{minipage}{0.95\textwidth}
\begin{center}
\caption{Properties of the simulations at $z=0$}
\label{tab:properties}
\resizebox{\textwidth}{!}{
\begin{tabular}{lllllllllllllllll}
\hline

(1) & (2) & (3) & (4) & (5) & (6) & (7) & (8) & (9) & (10) & (11) & (12) & (13) & (14) & (15) & (16) & (17)\\ 
Name		&	$M_\mathrm{*}$			&	$M_\mathrm{HI}$		&	$M_\mathrm{halo}$ 	& $M_{0.3}$ &	$M_\mathrm{half}$ 		& $M_V$	& $B-V$ & $V-I$ 	& $\mu_{0,V}$ 	& $R_e$ 	&	$R_{30,V}$	& $\mathrm{SFR}$	& $v_c$		& $\sigma_*$ &	$\mathrm{\langle[Fe/H]\rangle_{RGB}}$	& $12+\log(\mathrm{O/H})$\\
		&	$10^{6}\unit{M}_\odot$	&	$10^{6}\unit{M}_\odot$	&	$10^{9}\unit{M}_\odot$	& $10^7\unit{M_\odot}$	& $10^7\unit{M_\odot}$	& $\unit{mag}$ & $\unit{mag}$ & $\unit{mag}$ & $\unit{mag/arcsec^2}$ 	& $\unit{kpc}$ & $\unit{kpc}$	& $10^{-4}\unit{M_\odot/yr}$ & $\unit{km/s}$ & $\unit{km/s}$ & $\mathrm{dex}$ & $\mathrm{dex}$\\ \hline

DG9e9	& 0.10	& 1.14	& 4.94	& 2.65	& 0.77	& -8.70	& 0.31	& 0.80	& 24.52	& 0.14	& 0.43	& 1.07	& 14.54	& 10.9	& -1.45	& 7.22\\
DG10e9	& 1.89	& 1.38	& 5.38	& 1.66	& 3.46	& -10.98	& 0.54	& 1.00	& 24.60	& 0.43	& 1.26	& 1.08	& 12.57	& 13.0	& -1.46	& 7.94\\
DG12e9	& 0.15	& 0.26	& 6.33	& 1.98	& 0.71	& -8.29	& 0.55	& 1.00	& 25.46	& 0.18	& 0.45	& 0.14	& 11.92	& 12.0	& -1.40	& 7.71\\
DG13e9a	& 0.25	& 3.68	& 9.53	& 1.39	& 0.84	& -9.09	& 0.47	& 0.93	& 25.29	& 0.24	& 0.63	& 0.43	& 19.72	& 7.8	& -1.53	& 7.08\\
DG13e9b	& 1.23	& 3.07	& 5.93	& 0.58	& 2.06	& -10.50	& 0.60	& 1.01	& 25.41	& 0.49	& 1.24	& 0.44	& 15.15	& 10.3	& -1.66	& 7.74\\
DG15e9a	& 3.31	& 1.73	& 9.75	& 3.52	& 2.16	& -11.77	& 0.49	& 0.96	& 22.52	& 0.23	& 0.96	& 3.90	& 20.59	& 15.6	& -1.23	& 8.16\\
DG15e9b	& 4.41	& 26.35	& 9.22	& 1.20	& 2.67	& -12.40	& 0.41	& 0.89	& 23.18	& 0.42	& 1.59	& 10.82	& 24.03	& 12.9	& -1.44	& 7.64\\
DG20e9	& 3.74	& 4.05	& 12.83	& 1.18	& 2.51	& -11.80	& 0.52	& 0.99	& 23.68	& 0.40	& 1.40	& 4.32	& 14.73	& 12.3	& -1.40	& 8.03\\
DG50e9	& 14.83	& 66.40	& 34.70	& 0.58	& 5.97	& -13.62	& 0.43	& 0.91	& 23.03	& 0.69	& 2.65	& 49.06	& 24.24	& 15.3	& -1.48	& 7.93\\
DG1e11	& 1187.60	& 462.08	& 63.61	& 3.27	& 53.94	& -18.17	& 0.47	& 0.96	& 19.32	& 1.02	& 5.99	& 1425.77	& 60.35	& 34.9	& -0.91	& 8.76\\
\hline
DG15e9b-CT	& 8.88	& 31.57	& 8.64	& 0.48	& 4.42	& -12.98	& 0.45	& 0.93	& 23.64	& 0.69	& 2.39	& 11.12	& 19.92	& 11.6	& -1.31	& 7.81\\
\hline
DG10e9-NP3	& 20.43	& 41.47	& 7.04	& 0.24	& 10.54	& -13.57	& 0.55	& 1.00	& 24.32	& 1.23	& 3.83	& 19.14	& 19.20	& 14.5	& -1.60	& 8.11\\
DG12e9-NP3	& 11.85	& 14.14	& 6.37	& 0.23	& 9.41	& -12.94	& 0.62	& 1.01	& 25.01	& 1.26	& 3.46	& 5.62	& 15.16	& 14.2	& -2.05	& 8.08\\
DG13e9b-NP3	& 31.02	& 35.67	& 7.00	& 0.19	& 12.63	& -14.00	& 0.60	& 1.01	& 23.89	& 1.23	& 4.12	& 20.22	& 19.59	& 15.3	& -1.94	& 8.23\\
\hline
\end{tabular}}

\end{center}

\begin{tablenotes}
\item[] \textbf{Notes.} (1): the name of the simulation, (2): the stellar mass $M_\star$, (3): neutral gas mass $M_\mathrm{HI}$, (4): the virial mass of the dark matter halo $M_\mathrm{halo}$, (5): $M_{0.3}$, the total mass within $300 \unit{pc}$, (6): $M_\mathrm{half}$, the total mass within the half-light radius, (7): absolute V-band magnitude $M_V$, (8): $B-V$ colour, (9): $V-I$ colour, (10): central surface brightness brightness in the V-band $\mu_{0,V}$, (11): half-light radius $R_e$, (12): $R_{30, V}$, the radius where the fitted surface brightness profile reaches $30 \unit{mag/arcsec^2}$, (13): star formation rate, (14): circular velocity $v_{c}$, (15): stellar velocity dispersion $\sigma_*$, (16): average stellar iron abundance $\mathrm{\langle[Fe/H]\rangle_{RGB}}$ and (17) the oxygen-abundance of the ionized gas $12+\log(\mathrm{O/H})$.
\end{tablenotes}
\end{minipage}

\end{table*}
\begin{figure}[t!]
\includegraphics[width=0.45\textwidth]{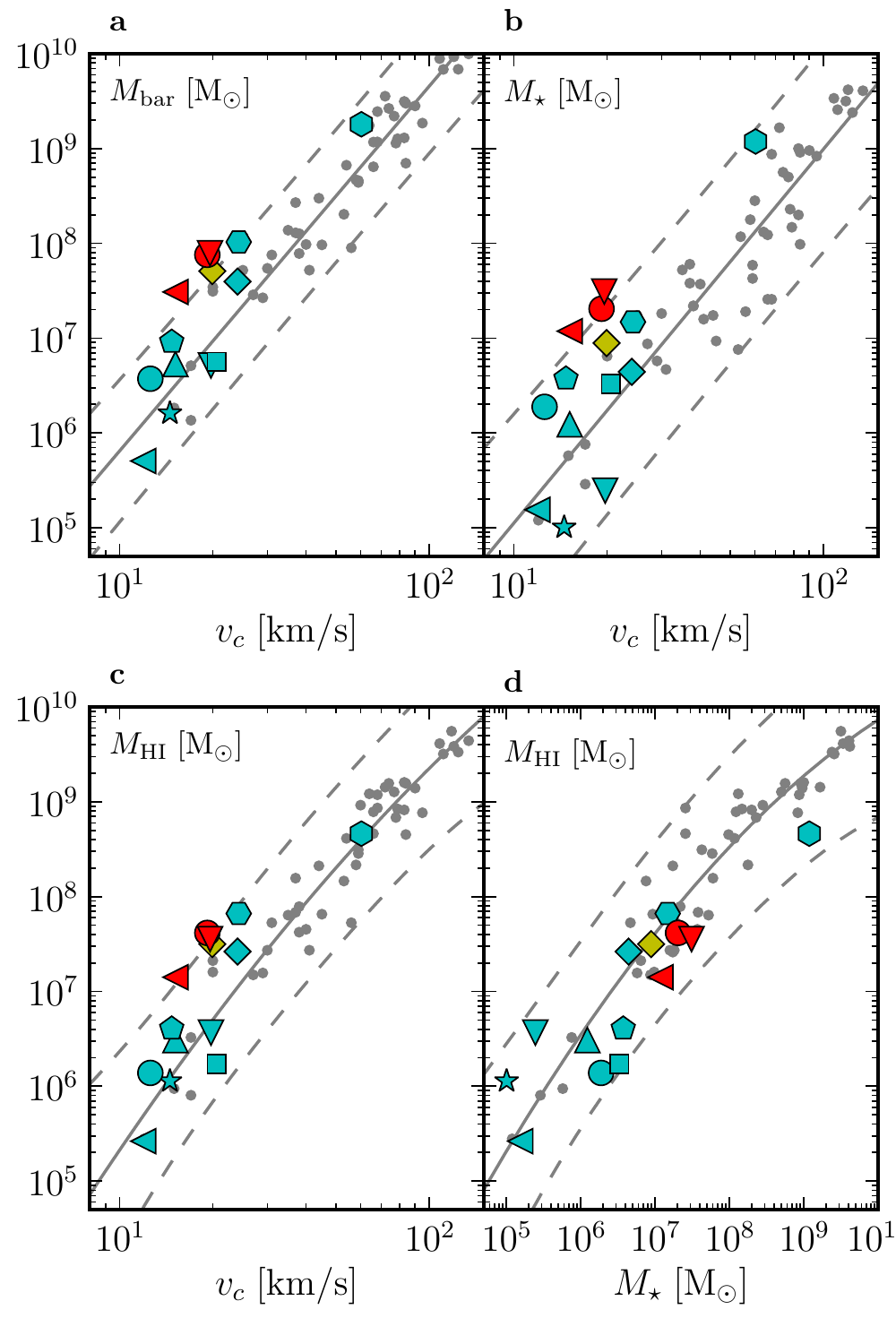}
\caption{The baryonic Tully-Fisher relations (a, b, and c) and stellar mass versus {\HI} mass (d). 
Our simulations compared with the observations \citep{mcgaugh12,
    bernsteincooper14}. 
The different symbols represent different simulations, as indicated in Table
 \ref{tab:simulations}. Blue, red and yellows symbols are simulations with {\popiii} feedback, without {\popiii} feedback and the convergence test respectively, while gray circles are observations.
The solid lines are the regression lines
  through the data (linear in a and b, quadratic in c
  and d) and the dashed lines delimit the $3\sigma$ prediction
  interval.}
\label{fig:btf}
\end{figure}

\subsection{Baryonic Tully-Fisher relation}

The baryonic Tully-Fisher relation (BTFR) relates the total
baryonic mass to the circular velocity \citep{mcgaugh12}. Since the inner mass density
profile is strongly influenced by the gravitational coupling between
gas and dark matter \citep{cloetosselaer12}, a galaxy's maximum circular
velocity depends non-trivially on the total mass, including dark
matter, and its star-formation history. The
simulations including {\popiii} feedback yield dwarf galaxies that lie
on the observed BTFR, in terms of total baryonic matter (Fig. \ref{fig:btf}a) as
well as the separate stellar and neutral gas component (Fig. \ref{fig:btf}b and \ref{fig:btf}c,
respectively). We confirm that these simulations reproduce the
observed relation between stellar mass and neutral gas mass as well
(Fig. \ref{fig:btf}d). The control simulations without {\popiii} 
feedback fall outside the $3\sigma$ prediction interval of the regression lines
fitted to the $v_c - M_{\rm bar}$ and $v_c - M_{\star}$ observations:
they produce $\sim 10$ times too many stars for their circular
velocity. In short, our simulations including {\popiii} feedback
succeed in creating faint, gas-dominated dwarf galaxies lying on the
BTFR, such as Leo~T, Leo~P, and Pisces~A, while the control
simulations without {\popiii} feedback do not.

\begin{figure}[t!]
\includegraphics[width=0.45\textwidth]{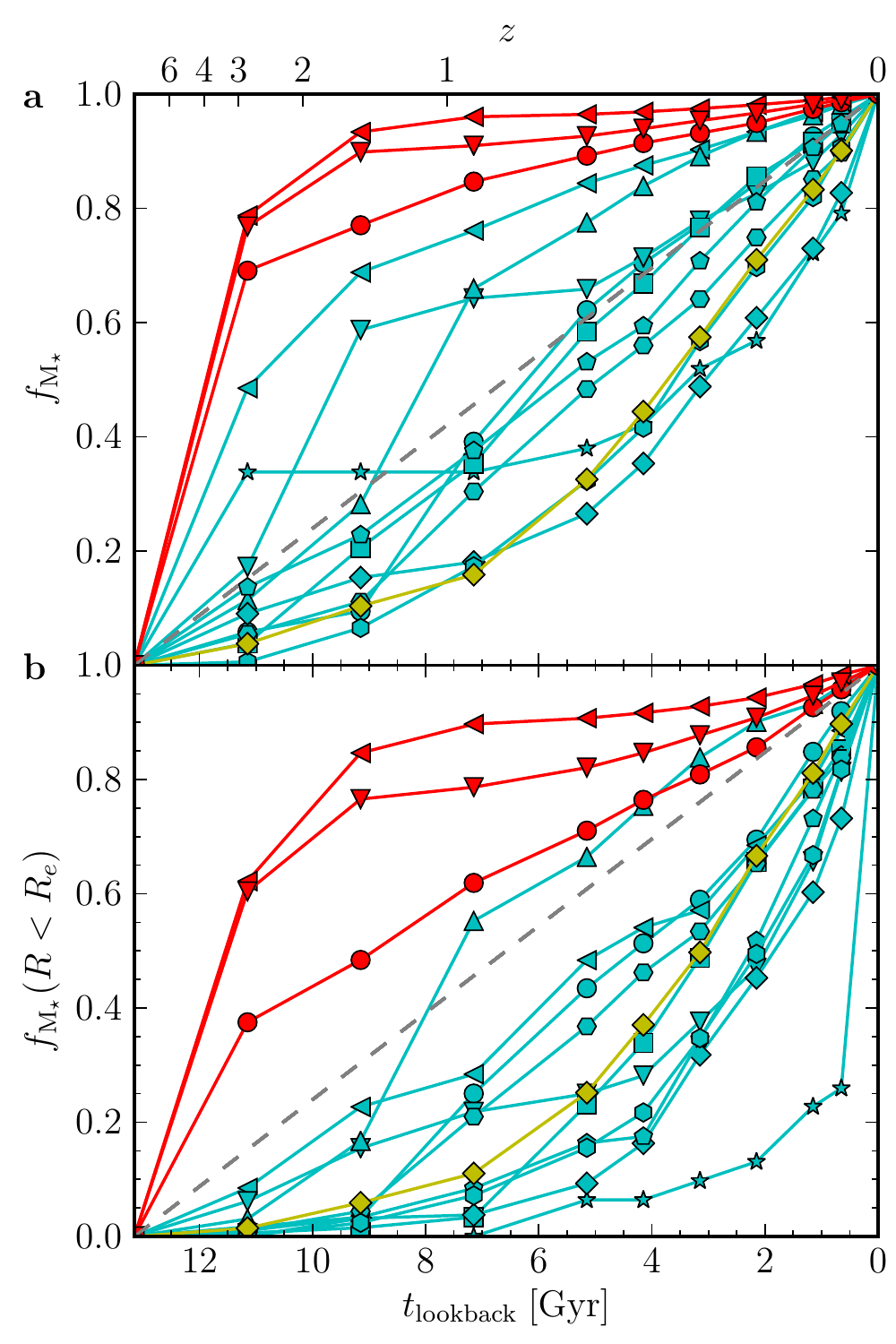}
\caption{Cumulative star formation histories. The fraction of the
  stars formed by a certain lookback time is shown, derived from all
  stars (a) within the simulation, and (b) within $1~R_{e}$. Symbols and colors are the same as in Fig. \ref{fig:btf}.
  The dotted lines shows the case for a constant
  star-formation rate.}
\label{fig:sfhs}
\end{figure}

\subsection{Cumulative star formation histories}

In Fig. \ref{fig:sfhs}, we present the CSFHs.
The upper panel (Fig. \ref{fig:sfhs}a) shows the CSFH of all the stars ever
formed in the simulation. 
%
However, when we only consider the
stars currently within a smaller projected distance from the galaxy
center, the stellar mass assembly appears significantly more
delayed. In Fig. \ref{fig:sfhs}b, we present the CSFH of the stars currently
within $1~ R_{e}$, which is more in line with
the area covered by current stellar-populations
surveys \citep{weisz14a}. These more central CSFHs look strikingly like
those of observed star-forming dwarf galaxies with comparable stellar
masses in and near the Local Group \citep{weisz14a}. In contrast,
previous simulations and semi-analytical models generally form too many stars early on \citep{weinmann12}.

The cause for the radial dependence of the CSFH is that later stellar generations tend to be
born within a more centrally concentrated volume than the earliest
stars and that stars can migrate radially and become unbound and
dispersed by the tidal forces accompanying merger events. Thus, older
stars, which are subjected to more merger events than younger ones,
are preferentially affected. In the turbulent ISM of the simulations
with {\popiii} feedback, the first stars show significant outwards
radial migration and are subsequently strongly affected by tidal
forces. This is much less the case in the control simulations without
{\popiii} feedback. As a consequence, these have much less radially
varying CSFHs. 

\begin{figure}[t!]
\includegraphics[width=0.45\textwidth]{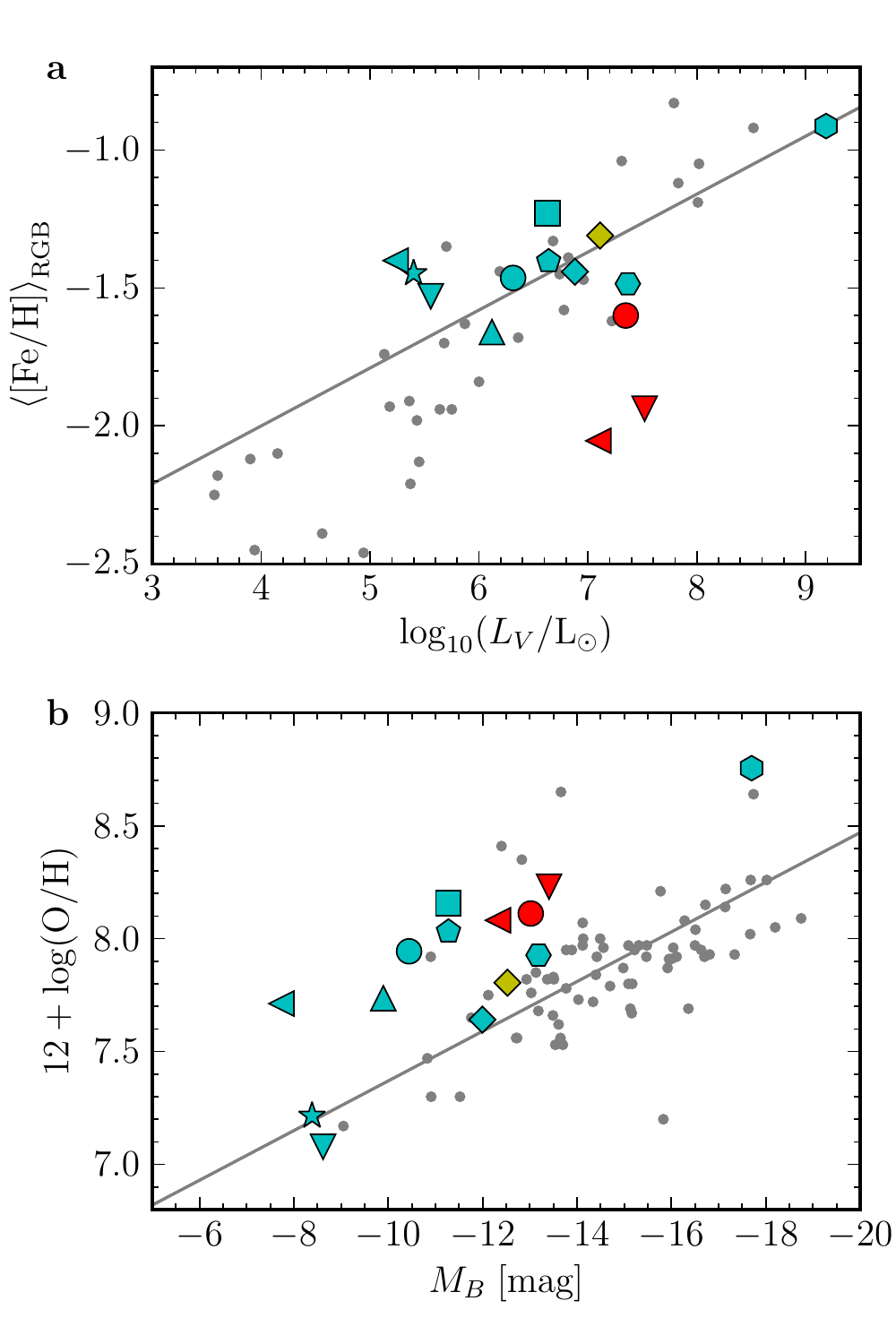}
\caption{Metallicities. (a) The stellar Iron abundance and (b) the
  Oxygen abundance of the ionized gas around star forming regions of
  the simulations compared with observations of dwarf
  irregulars \citep{croxall09, berg12, skillman13,
    kirby13}. Symbols and colors are the same as in Fig. \ref{fig:btf}. The fit to the data in (a) is
  $\langle\mathrm{[Fe/H]}\rangle_\mathrm{RGB} = -2.84 +
  0.21\log_{10}(L_V/\unit{L_\odot})$ \citep{kirby13} and in (b)
  $12+\log(\mathrm{O/H)} = 6.27-0.11M_B/\unit{mag}$ \citep{berg12}.}
\label{fig:metal}
\end{figure}

\subsection{Metallicities}
\label{sec:metals}
Metallicity measurements for nearby dwarf galaxies are often based on
spectroscopic observations of bright stars on the Red Giant Branch
(RGB) \citep[e.g.][]{starkenburg10}. The number of {\popii} RGB stars
within the half-light radius of the simulated dwarfs ranges from $\sim
65$ for the faintest one up to $\sim 2000$ for the brightest
ones. The estimated number of {\popiii}
RGB stars is zero in all simulated dwarfs presented
here. The mean stellar iron abundance of stars that
are now on the RGB in the simulated dwarfs agrees very well with the
observations (Fig. \ref{fig:metal}a). The oxygen abundance of the ionized gas
surrounding star-forming regions is somewhat on the high side but also
in broad agreement with the observations (Fig. \ref{fig:metal}b). Oxygen is an
element forged mostly by core-collapse supernovae while iron is formed
abundantly in \snia. The simulations without
{\popiii} feedback form most of their stars early on and therefore
have an inordinately large low-metallicity RGB population and hence
fall significantly below the observed luminosity-metallicity relation of
star-forming dwarf galaxies. 

\subsection{Metallicity distribution functions}

\begin{figure}[t!]
\includegraphics[width=0.45\textwidth]{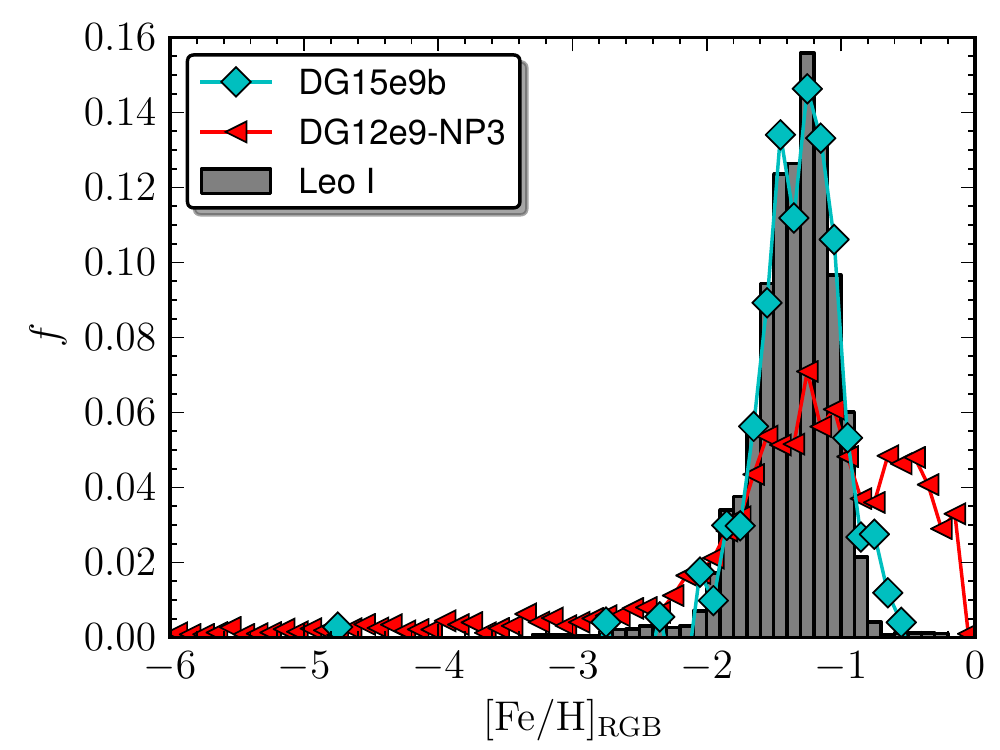}
\caption{Metallicity distribution function. The MDFs of a simulation without \popiii~ feedback and of one including it, compared to the MDF of Leo I.}
\label{fig:mdf}
\end{figure}

In the previous section, we discussed the global metallicity of the simulated galaxies. However, metallicity distribution functions (MDFs) are available in the literature for many dwarf galaxies \citep[e.g.][]{kirby13}.
To construct MDFs from our simulations, we again limit ourselves to the stars inside $1~R_e$ and take the predicted number
of RGB stars within each stellar population into account. Fig. \ref{fig:mdf} shows the MDF of Leo I, taken from \cite{kirby13}, along with the MDFs of DG15e9b and DG12e9-NP3, which both have stellar masses similar to Leo I. 
The MDFs of DG12e9-NP3 and the other simulations without \popiii~ feedback all have a very long low-metallicity tail: $\sim 1\%$ of the RGB stars is in the form of stars with $\mathrm{[Fe/H]} < -5$. This long tail is absent in observations and such metal-poor stars would be more easily detected. On the other hand, the low-metallicity tail is not present in DG15e9b and the other simulations including \popiii~ feedback and the MDF of DG15e9b shows very good agreement with that of Leo I. The simulation has slightly more metal-rich stars which could be because it is still actively star-forming, while Leo I has recently been quenched \citep{weisz14a}.

This gives further evidence that the first generation of stars did indeed have very different properties than the stars we see today: if such extremely metal-poor stars would have an IMF similar to \popii~ stars, they should be detectable. However to date, no such stars have been found in dwarfs \citep{starkenburg10}.

\subsection{$M_\star-M_\mathrm{halo}$-relation}

\begin{figure}[t!]
\includegraphics[width=0.45\textwidth]{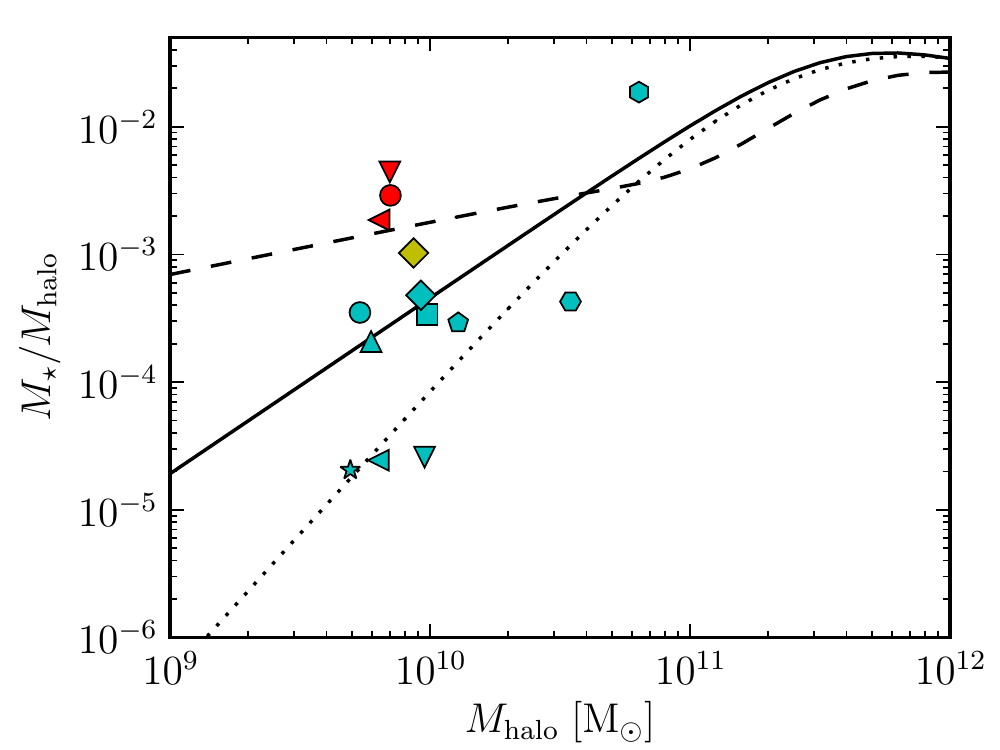}
\caption{$M_\star-M_\mathrm{halo}$-relation. The simulations compared to the abundance matching relations of \cite{guo10} (dotted line), \cite{moster13} (solid line) and \cite{behroozi13} (dashed line). Symbols and colors are the same as in Fig. \ref{fig:btf}.}
\label{fig:mstarmhalo}
\end{figure}

The $M_\star-M_\mathrm{halo}$-relation relates the stellar mass to the dark matter 
mass and is typically obtained using the abundance matching technique; it is observationally less tractable but theoretically more
straightforwardly calculated. We find that the models with {\popiii}
feedback lie within the range predicted by abundance matching
techniques \citep[][respectively the dotted,  solid and dashed line in Fig. \ref{fig:mstarmhalo}]{guo10, moster13, behroozi13}. Note that
the simulations with the largest circular velocity do not necessarily
have the most massive dark matter halos since the former also depends
on the spatial distribution of the matter. At a given halo mass, the
scatter on the stellar mass is considerable. This suggests that the
abundance matching approach likely looses its applicability in this
mass regime. The control simulations without {\popiii} feedback lie
significantly above the predicted $M_\star-M_\mathrm{halo}$-relations.

\subsection{\HI~ distribution}

\begin{figure*}[t!]
\begin{minipage}{\textwidth}
\begin{center}

\includegraphics[width=\textwidth]{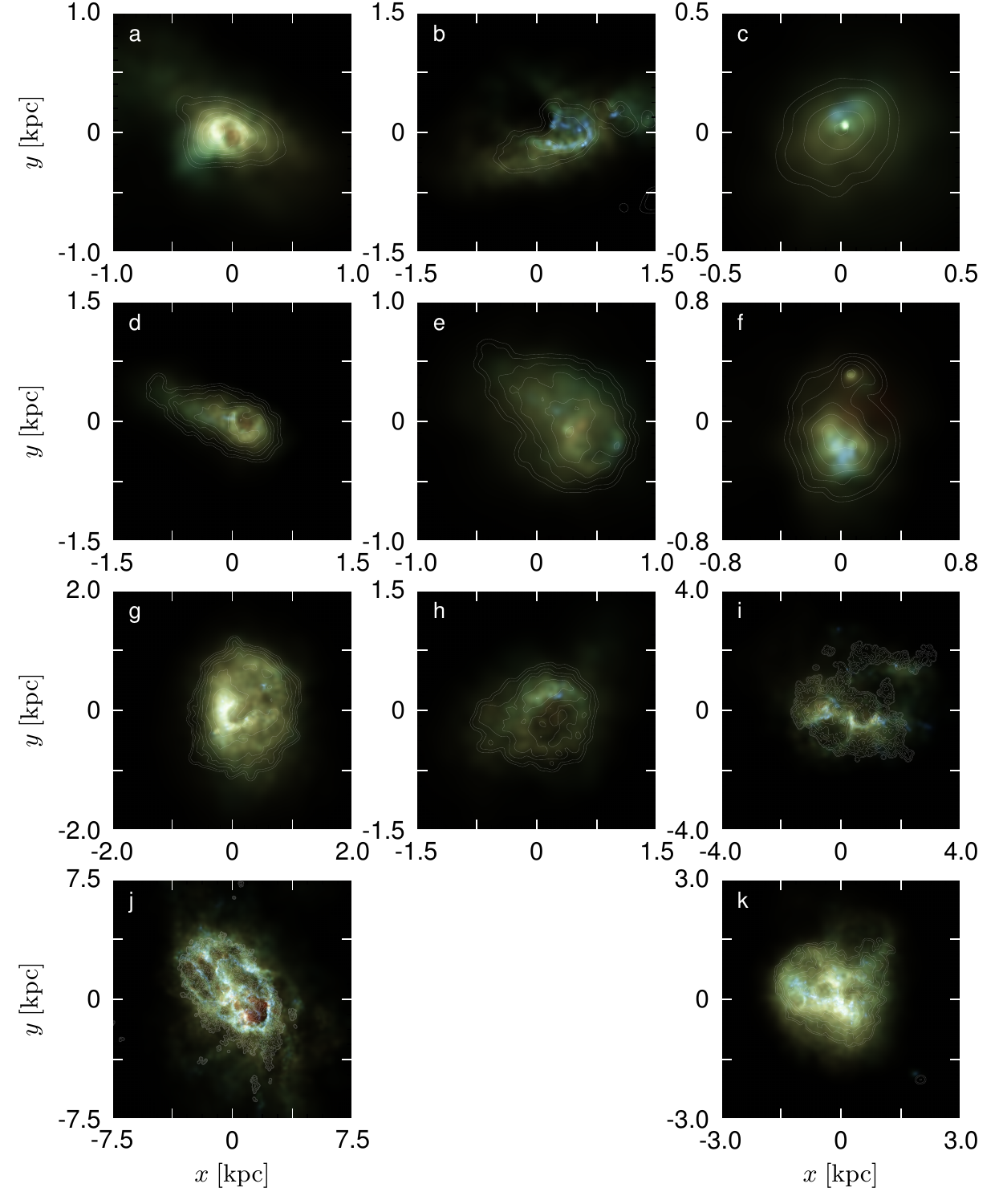}
\end{center}

\end{minipage}

\caption{\HI~ distribution of the simulations. The gas is rendered using the publicly available ray tracing software Splotch, overplotted with the \HI~ density isophotes, with the lowest level $N_\mathrm{HI} =
10^{19} \unit{cm^{−2}}$ with each next level an increment of a factor 4. The beam size for the \HI~ contours is $40 \unit{pc}$. (a) DG9e9, (b) DG10e9, (c) DG12e9, (d) DG13e9L, (e) DG13e9E, (f) DG15e9L, (g) DG15e9E, (h) DG20e9,  (i) DG50e9, (j) DG1e11 and (k) DG15e9b-CT.
}

\label{fig:HIMaps}
\end{figure*}

Fig. \ref{fig:HIMaps} shows the \HI~ distribution, rendered with the publicly available ray tracing package Splotch\footnote{http://www.mpa-garching.mpg.de/ kdolag/Splotch}, overplotted with their \HI~ density contours, with beam sizes of $40 \unit{pc}$.
While this shows that our simulations look qualitatively like real dwarf galaxies, we can look at them more quantitatively.

\begin{figure}[t!]
\includegraphics[width=0.45\textwidth]{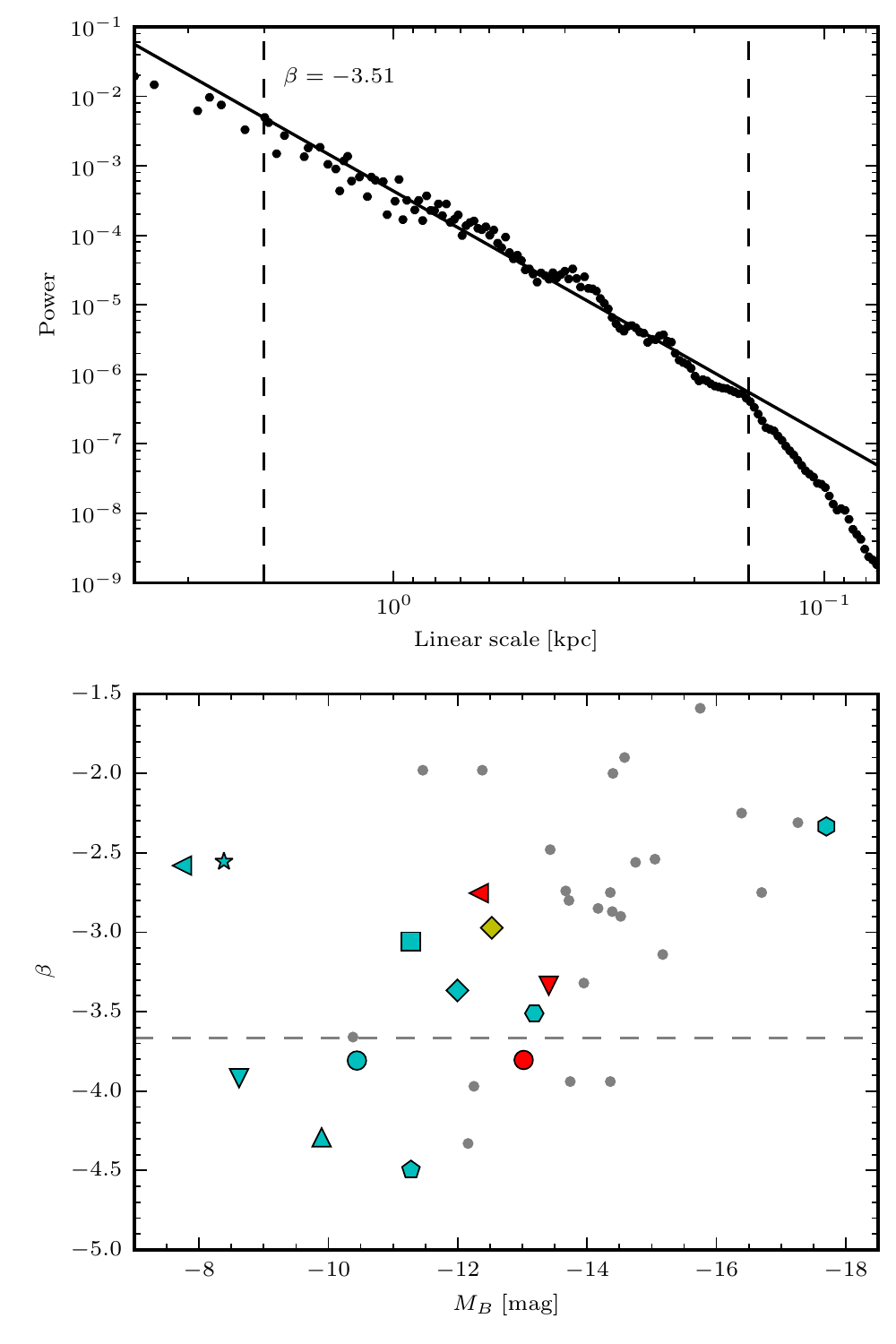}
\caption{Spatial power spectra of the \HI. (a) The power spectrum of
the spatial distribution of HI in DG50e9. The solid grey line shows the power-law fit while
the vertical dashed lines show the fitting range. (b) Spectral indices of the power spectra
of our simulations, compared with dwarfs in the Little Things survey\citep{zhang12}. The horizontal
dashed line shows the value of 3D Kolmogorov-type turbulence. Symbols and colors are the same as in Fig. \ref{fig:btf}.}
\label{fig:powerspectrum}
\end{figure}

The substructure in the ISM can be quantified using Fourier transform power spectra of
the {\HI} maps. These can be compared with radio
observations of dwarf galaxies via the spectral index $\beta$ of the
power spectrum $P(k) \propto k^\beta$, where $k$ is the wave
number (Fig. \ref{fig:powerspectrum}). Like for most real dwarfs in this luminosity regime, this
index scatters around $\beta=-11/3\approx -3.67$, the value expected
in the case of a three-dimensional Kolmogorov-type incompressible,
subsonic turbulence \citep{zhang12}. This constitutes a test of the feedback
prescription, responsible for blowing ``holes'' in the neutral gas and
keeping the ISM thick and turbulent. 
\newline

\subsection{Other properties}

\begin{figure*}[t!]
\begin{minipage}{\textwidth}
\begin{center}

\includegraphics[width=\textwidth]{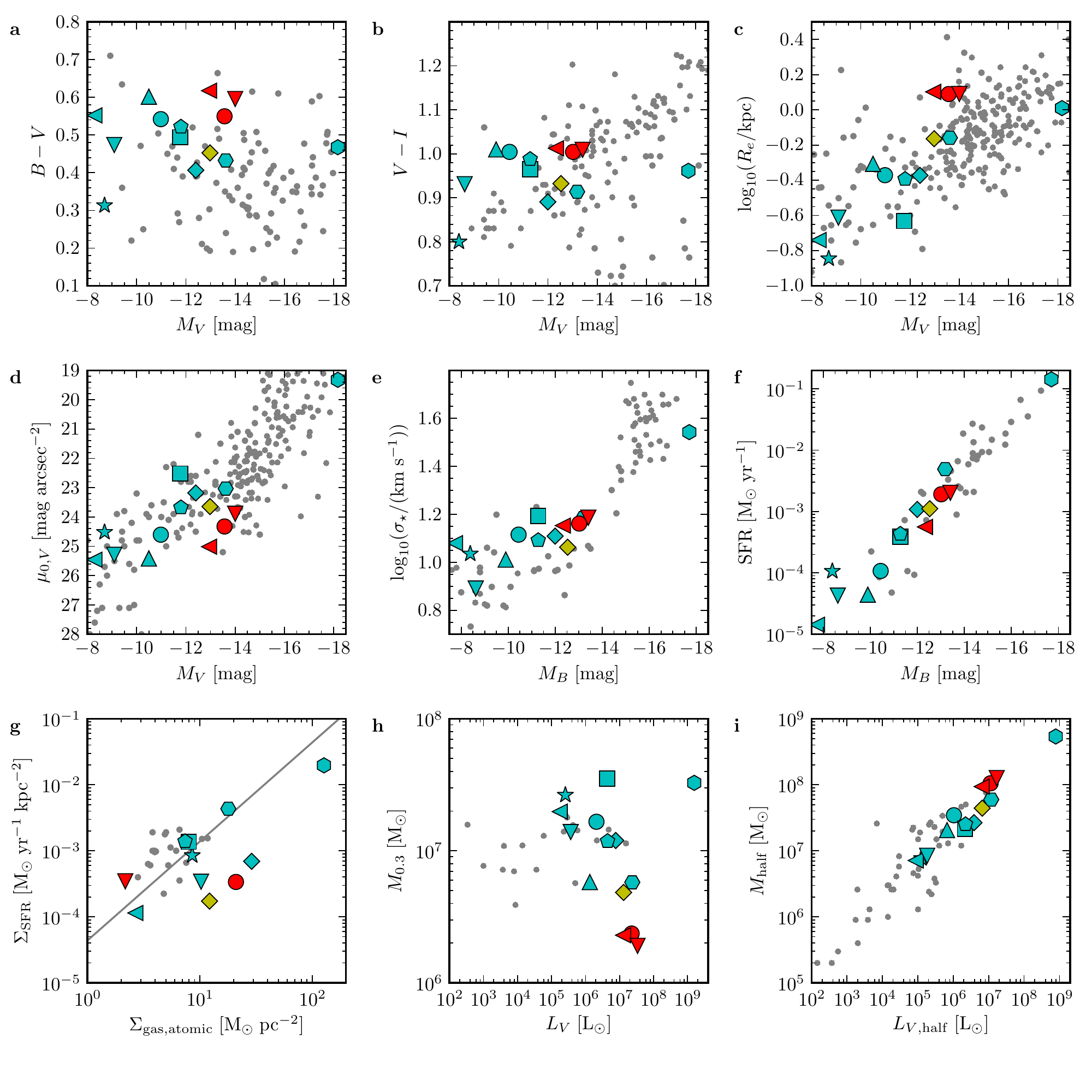}
\end{center}

\end{minipage}

\caption{Other scaling relations. (a) $B-V$ colour, (b) $V-I$ colour, 
  (c) Half light radius $R_e$, (d) Central
  surface brightness in the V-band $\mu_{0, V}$, (e) Stellar velocity dispersion $\sigma_\star$,
  (f) Star formation rate, (g) Kennicutt-Schmidt relation, (h) Total mass within $300 \unit{pc}$, (i) Total mass within $1\ R_e$ compared to observations \citep[in grey;][]{graham03, hunter06, strigari08, vennik08,
    derijcke09, makarova09, mcconnachie12, rhode13, mcquinn13, karachentsev13, collins14, hopp14,
    tollerud15, roychowdhury15}. Symbols and colors are the same as in Fig. \ref{fig:btf}.}

\label{fig:properties}
\end{figure*}

We also compare the optical colours, half-light radius, the central 
surface brightness, the central stellar velocity dispersion, the star-formation
rate, the Kennicutt-Schmidt relation and the total mass within $300\ \unit{pc}$ and $1\ R_e$ of our models with the observations (Fig. \ref{fig:properties}a$-$i) . We
generally find good agreement between the observed scaling relations
and our simulations with {\popiii} feedback.
The control simulations without {\popiii} feedback are
generally more extended and have lower central stellar and mass
densities than observed dwarfs.

We note that we reproduce both $M_{0.3}$, the observed total mass within $300\ \unit{pc}$ \citep[][Fig. \ref{fig:properties}h]{strigari08}, and $M_\mathrm{half}$, the total mass within the half-light radius \citep[][Fig. \ref{fig:properties}i]{collins14}. Furthermore, although these relations where determined for dSphs, with $L_V \lesssim 10^7\unit{L_\odot}$, our simulations agree with them over the entire luminosity range, and they predict an extension of this relation to $L_V \lesssim 10^9\unit{L_\odot}$. For some simulations, $M_{0.3}$ seems to be on the high side however. This is further discussed in the next section.

\subsection{Cusp versus core}

\begin{figure}[t!]
\includegraphics[width=0.45\textwidth]{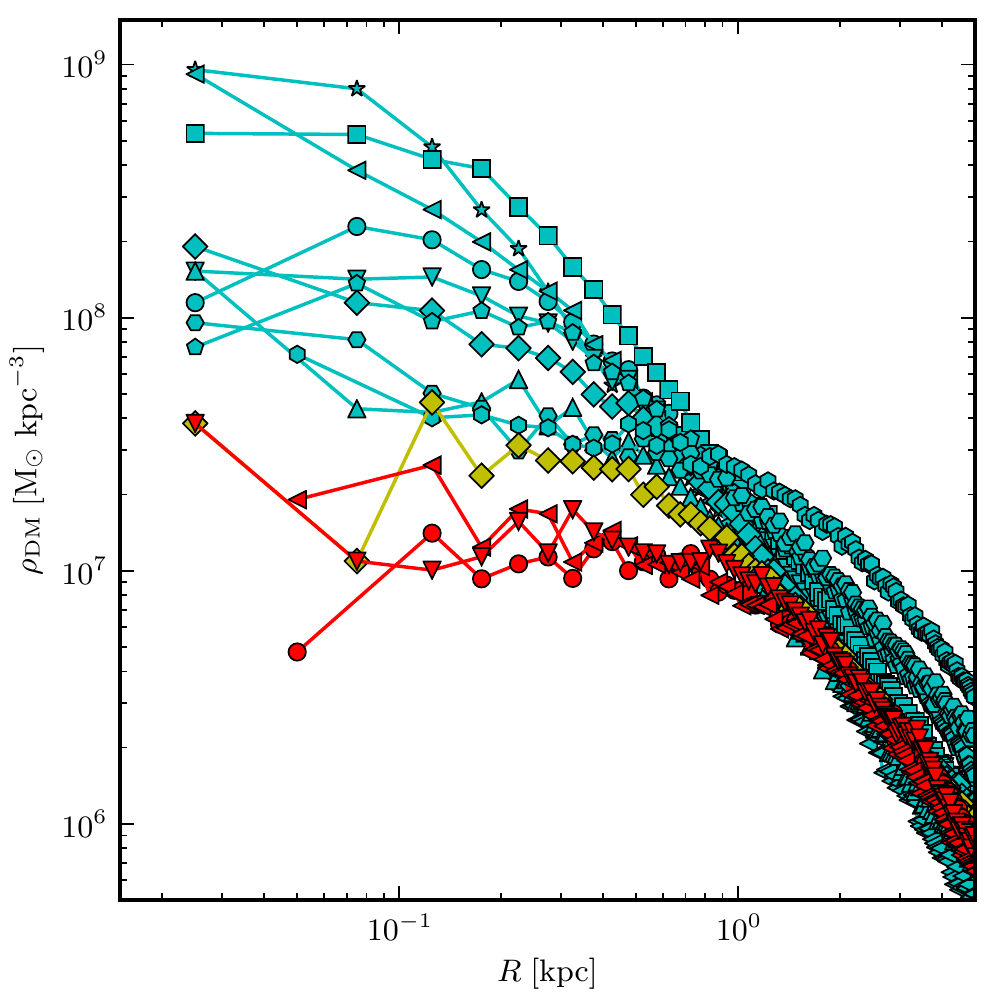}
\caption{Dark matter density profiles of the simulations. Symbols and colours are as in Figure \ref{fig:btf}.}
\label{fig:DMdensity}
\end{figure}

Cosmological simulations including only gravity predict a universal density profile with a cusp in the central regions \citep{navarro96}. On the other hand, observations generally favor a cored density profile \citep{deblok08, salucci12}. It has been shown that including baryonic effects in simulations can drastically affect the dark matter density profiles and lead to cored density profiles \citep{read05, governato10, cloetosselaer12}.

Fig. \ref{fig:DMdensity} shows the dark matter density profiles of the simulations. The models can be divided into three groups: DG9e9, DG12e9 and DG15e9a (group 1) have a high dark matter density in the central regions ($\rho_\mathrm{DM,0} \sim 10^9\unit{M_\odot\ kpc^{-3}}$); group 2 consists of the models without \popiii\ feedback (DG10e9-NP3, DG12e9-NP3 and DG13e9b-NP3, in red), which have a very low central dark matter density ($\rho_\mathrm{DM,0} \sim 10^7\unit{M_\odot\ kpc^{-3}}$); the central densities for the remaining models (group 3) lies in between these groups ($\rho_\mathrm{DM,0} \sim 10^8\unit{M_\odot\ kpc^{-3}}$). We can now have a look to see if these groups manifest themselves in different scaling relations as well. From panels a and c in Fig. \ref{fig:btf}, one might argue that group 1 lies lower on the BTFR, in terms of $M_\mathrm{bar}$ and $M_\mathrm{HI}$, although only marginally. However, this is not true in terms of $M_\star$ (panel b). Group 3 lies significantly above the BTFR, indicating that they have unrealistic shallow inner density profiles.

Group 3 also has very early SFHs, although this will more likely be the cause of the shallow density profiles, rather than the other way around. Group 1 and group 2 can not be distinguished from their SFHs.

Group 1 lies higher in the stellar metallicity plot (Fig. \ref{fig:metal}a). One possible explanation is that because of their higher central densities, they can retain their metals more effectively. However, this is not necessarily the only explanation, since the division in groups based on their density profiles was done at $z=0$ and it is not clear how much these density profiles change over time.

There is no clear distinction between group 1 and 2 in the $M_\star-M_\mathrm{halo}$-relation (Fig \ref{fig:mstarmhalo}), indictating that the inner dark matter density profile and the total halo mass influence galaxy properties in different ways.

Not suprisingly, group 1 clearly has smaller $R_e$ and $\mu_{0,V}$ (Fig. \ref{fig:properties}c-d), while group 3 has larger $R_e$ and $\mu_{0,V}$.  Group 1 and group 3 also have larger, respectively smaller, $M_{0.3}$ (Fig. \ref{fig:properties}h).
DG1e11, which was categorized in group 2, also has a small $R_e$ and large $M_{0.3}$, but this is mostly because of its high concentration of stars in its central region, rather than its dark matter properties. Group 1 seems to have a higher stellar velocity dispersion $\sigma_\star$, although only very slightly, since $\sigma_\star$ was determined at $1\ R_e$, so at smaller radii for group 1. Because of the interplay between $R_e$ and central density, all groups behave in the same way in terms of $M_\mathrm{half}$ (Fig. \ref{fig:properties}i).

All of this indicates that to identify galaxies with a cusped dark matter density profile, the most likely candidates are the ones with small half-light radii and high total masses within a small, fixed radius. Furthermore, from our simulations it seems unlikely that any dSph galaxies analyzed in \cite{strigari08} have a cusped density profile.

\subsection{\popiii~ to \popii~ transition}

\begin{figure}[t!]
\includegraphics[width=0.45\textwidth]{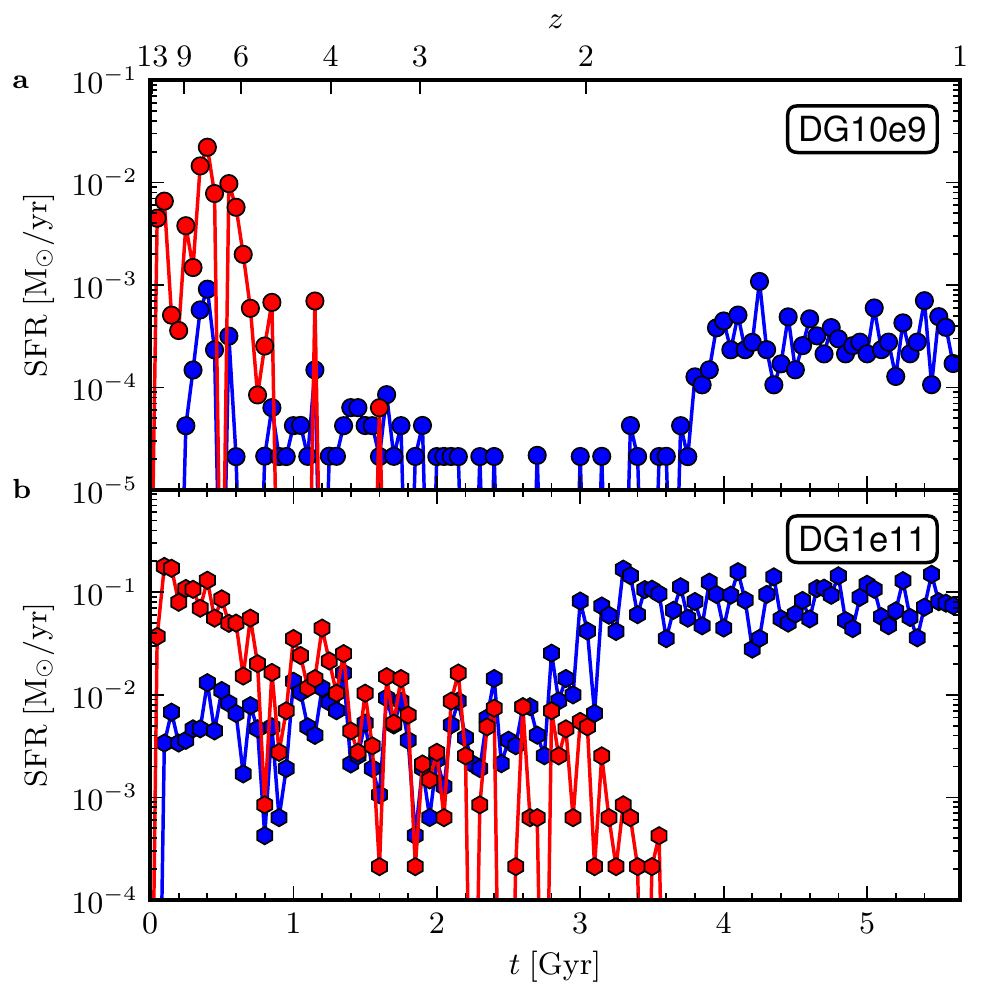}
\caption{Transition of \popiii~ to \popii~ star formation. Star formation rate of the \popiii~ stellar populations (red) and of the \popii~ populations (blue) of DG10e9 (a) and DG1e11 (b), one of our least and our most massive models, respectively. All stellar particles formed in the simulations are taken into account.}
\label{fig:sfrpop3}
\end{figure}

The transition of \popiii~ to \popii~ star formation occurs, in terms of metallicity, at $\mathrm{[Fe/H]} = -5$ in our models. However, it is not immediately clear at what time this transition occurs and how abrupt it is. Fig. \ref{fig:sfrpop3} shows the star formation rate of \popiii~ stars and of \popii~ stars (in red and blue, respectively) for one of our least massive massive (DG10e9, top panel) and our most massive model (DG1e11, bottom panel). The reason we show DG10e9 rather than DG9e9, is because there is no star formation in this model for $3 \lesssim z \lesssim 1$, as seen in Fig. \ref{fig:sfhs}. Before 1 Gyr, the \popiii~ stars dominated the overall star formation, although there is already some low level \popii~ star formation this early on. This \popiii~ star formation declines in time, after which the \popii~ star formation starts dominating the global star formation and after some time, all stars are being formed out of sufficiently enriched gas. The exact time of the last \popiii~ stars being formed seems to differ significantly between the two models shown in Fig. \ref{fig:sfrpop3}: the transition occurs much earlier for the least massive model, at $t\approx 1.6 \unit{Gyr}$, while the most massive model still has \popiii~ star formation until $t\approx 3.5 \unit{Gyr}$. This trend for later transition times for more massive galaxies is true for all our simulations.
\newline

Although this gives a general idea of how and when this transition between star formation modes might have occurred in the universe, we do not wish to overinterpret these results, since the cut-off metallicity is not well constrained and the 
transition will most likely have been more smoothly than the sharp bimodal model we obtained in our models. However, it does serve as an important check and we confirm that we do not have any \popiii~ star formation at low redshift, which would be more easily detected observationally.

\section{SUMMARY}

We have shown that, with the inclusion of the fierce UV
radiation emitted by {\popiii} stars in numerical simulations, we can 
reproduce the most important observed properties
of isolated gas-rich dwarf galaxies within the $\Lambda$CDM
paradigm. This energetic feedback from
the very first generation of stars is crucial
in suppressing the star-formation rate in a dwarf's progenitors and
thus delaying its star-formation history. We stress that no parameters
were fine-tuned to obtain these results.

Furthermore, the expected number of {\popiii} RGB stars is essentially zero, in agreement with the most metal-poor stars found in dwarf galaxies \citep{starkenburg10}.

Of course, there is still room for improvement. For instance, the Oxygen abundances predicted by our simulations are generally on the high side, as is the scatter between the values for the total mass within $300 \unit{pc}$. It remains to be seen if and how these remaining disagreements between models and data can be resolved.

In short, we have provided
numerical evidence that the first stars that formed in
the early Universe are essential for producing isolated,
gas-dominated dwarf galaxies in simulations with broadly
the same chemical, kinematical, and structural properties as real
dwarf systems in the nearby Universe. 

Finally, we wish to stress the
crucial importance of mimicking the observations as closely as
possible (e.g. by deriving RGB-weighted mean metallicities). 
This is the only meaningful way in which the validity of
numerical simulations can be tested.

\acknowledgments
R.V. thanks the Interuniversity Attraction Poles Programme
   initiated by the Belgian Science Policy Office (IAP P7/08
   CHARM). S.D.R. and B.V. thank the Ghent University Special Research
   Fund (BOF) for financial support. We thank the anonymous referee for his/her comments that improved the content and presentation of the paper. We thank Mina Koleva, Annelies
   Cloet-Osselaer, Gianfranco Gentille and Till Sawala for fruitful discussions. We thank Volker
   Springel for making the GADGET-2 simulation code publicly
   available.

\end{document}